\documentstyle[12pt]{article}
\title{Unparticle as a particle with arbitrary mass}
\author{Hrvoje Nikoli\'c \\
Theoretical Physics Division, Rudjer Bo\v{s}kovi\'{c} Institute, \\
P.O.B. 180, HR-10002 Zagreb, Croatia \\
{\normalsize hrvoje@thphys.irb.hr} \\
\makebox[1in]{} \\
}
\date{\today}
\begin{document}
\maketitle
\begin{abstract}
The unparticle field operator can be expanded in terms of creation and destruction
operators corresponding to particles with a continuous mass spectrum. Hence, when the 4-momentum of an unparticle is measured, then the
unparticle manifests as an ordinary particle with a definite (but arbitrary) mass. 
\end{abstract}
\vspace*{0.5cm}
PACS: 11.10.-z, 03.70.+k \newline
Keywords: Unparticle; particle \newline

\section{Introduction}

The possibility of the existence of a scale invariant sector of an 
effective field theory recently proposed in \cite{georgi} has received
a considerable attention. (For the most recent works with extensive
lists of references see, e.g., \cite{misc1,misc2,misc3}.)
It has been argued \cite{georgi} that physics of that sector cannot be
described in terms of particles, which is why this sector has been
dubbed {\it unparticle} stuff. In particular, it has been obtained \cite{georgi}
that the unparticle stuff with the scale dimension $d_{\cal U}$ 
may appear as a non-integer number $d_{\cal U}$ of invisible particles.

After the initial work \cite{georgi}, it has been realized that certain aspects of unparticle physics can be viewed as physics of quantum fields with a
continuous mass spectrum \cite{step,kras,gaete}. This suggests that, at the
fundamental level, the unparticle stuff may not be so different from the
ordinary notion of particles appearing in quantum field theory. 
In this paper we further explore such a particle interpretation of the unparticle stuff. The next section deals with particles with a discrete and continuous
mass spectra, while the relation with unparticles is discussed in Sec.~\ref{SEC3}. Additional remarks on canonical quantization of unparticle fields are given in Sec.~\ref{SEC4}.

\section{Particles with various masses}
\label{SEC2}

Consider a discrete collection of free hermitian scalar fields $\phi(x,m)$, 
each having a different mass $m$. These fields satisfy the canonical 
equal-time commutation relations
\begin{equation}\label{e0}
[\phi({\bf x},m),\pi({\bf x}',m')]=i\delta^3({\bf x}-{\bf x}')\delta_{mm'} ,
\end{equation}
where $\pi=\partial_0\phi$ are the canonical momenta. By expanding
the fields in terms of plane waves in the standard way, one obtains
the creation and destruction operators $a^{\dagger}({\bf k},m)$ and 
$a({\bf k},m)$, respectively, that obey the commutation relations
\begin{equation}\label{e1}
[ a({\bf k},m), a^{\dagger}({\bf k'},m') ] = f({\bf k},m) 
\delta^3( {\bf k}-{\bf k}' ) \delta_{mm'} ,
\end{equation}
with other commutators vanishing.
The function $f({\bf k},m)$ is a matter of normalization, with a few 
different normalizations appearing in the literature. 
As is well known, these commutation relations imply that the quantum fields
exhibit a particle interpretation; the representation space of the field algebra
is constructed by acting with the creation operators
$n=1,2,3,\ldots$ times on the vacuum $|0\rangle$, corresponding to states
with an {\em integer} $n$ number of particles. 
In particular, the most general 1-particle state is a state of the form
\begin{equation}\label{e2}
|1\rangle = \sum_m \int d^3k \, c({\bf k},m) |{\bf k},m\rangle ,
\end{equation}
where $|{\bf k},m\rangle = a^{\dagger}({\bf k},m) |0\rangle$
and $c({\bf k},m)$ is an arbitrary function normalized so that
$\sum_m \int d^3k \, |c({\bf k},m)|^2=1$.

The physical meaning of a superposition (\ref{e2}) involving various types
of particles specified by different masses $m$ may seem peculiar.
Nevertheless, such a peculiarity disappears when the mass $m$ is measured.
According to the standard rules of quantum mechanics, when an observable is
measured then the value of this observable attains a definite value.
Thus, when $m$ is measured, one observes a particle of a definite type
specified by the mass $m$. The probability that the measured mass will have the
value $m$ is equal to $\int d^3k \, |c({\bf k},m)|^2 $. Furthermore, if the
3-momentum is also measured simultaneously with the mass, one observes a particle
with a definite 4-momentum $k=({\bf k},k_0)$, where $k_0$ and $m$ are related
as
\begin{equation}\label{e3}
m^2=k_0^2-{\bf k}^2 \equiv k^2 .
\end{equation}

Now let us generalize
it to the case of a continuous mass spectrum. Clearly,
the Kronecker $\delta_{mm'}$ gets replaced by the Dirac function 
$\delta(m-m')$. However, to provide a manifest Lorentz covariance
in the momentum space,
it is more convenient to replace the independent variables $({\bf k},m)$
by another set of independent variables $({\bf k},k_0)\equiv k$, where
the variables $m$ and $k_0$ are related as in ({\ref{e3}).
Thus, with an appropriate choice of the normalization $f$, 
(\ref{e1}) generalizes to
\begin{equation}\label{e1'}
[ a(k), a^{\dagger}(k') ] =  \delta^4(k-k') . 
\end{equation}
Such a commutation relation has also been introduced in \cite{kras2}.
Clearly, just as (\ref{e1}), such a commutation relation also admits a particle interpretation;
an $n$-particle state is obtained by acting $n$ times with the 
creation operators $a^{\dagger}(k)$ on the vacuum $|0\rangle$.
In particular, (\ref{e2}) generalizes to  
\begin{equation}\label{e2'}
|1\rangle = \int d^4k \, c(k) |k\rangle ,
\end{equation}
where $|k\rangle = a^{\dagger}(k) |0\rangle$
and $c(k)$ is an arbitrary function normalized so that
$\int d^4k \, |c(k)|^2=1$.
Analogously to the case of a discrete mass spectrum discussed above, when the 4-momentum $k$ is measured, 
then one observes an ordinary particle with a definite 4-momentum
$k$. In this case the mass $m^2=k^2$ also takes a definite value. 
The only difference with respect to the discrete-mass case is the fact 
that now the mass may take {\em any} value $m$, with the probability density 
equal to
$\int d^4k \, \delta(|k|-m) |c(k)|^2$.

\section{The relation with unparticle fields}
\label{SEC3}

The physics of the particle creation and destruction operators obeying
(\ref{e1'})
can be viewed as physics of a one-parameter family of massive fields
$\phi(x,m)$, where $m$ is a continuous parameter. But there is also
a different view of the same creation and destruction operators. 
Instead of dealing with the family of fields $\phi(x,m)$, one can deal
with a single field
\begin{equation}\label{exp}
\Phi(x)=\int \frac{d^4k}{(2\pi)^4} F(k)\theta(k_0)\theta(k^2)
[a(k)e^{-ikx}+a^{\dagger}(k)e^{ikx}] .
\end{equation}
The function $F(k)$ can be determined by imposing the requirement of scale invariance.
Following \cite{georgi}, the requirement of scale invariance implies
that the 2-point function must have the form
\begin{equation}\label{2p1}
\langle 0|\Phi(x)\Phi(0) |0\rangle = A_{d_{\cal U}}
\int \frac{d^4k}{(2\pi)^4} (k^2)^{d_{\cal U}-2} \theta(k_0)\theta(k^2) e^{-ikx} .
\end{equation}
On the other hand, (\ref{e1'}) implies
\begin{equation}
\langle 0|a(k)a^{\dagger}(k') |0\rangle =\delta^4(k-k'),
\end{equation}
so, using the fact that the step function satisfies
$\theta^2=\theta$, the expansion (\ref{exp}) leads to
\begin{equation}\label{2p2}
\langle 0|\Phi(x)\Phi(0) |0\rangle = 
\int \frac{d^4k}{(2\pi)^4} \frac{F^2(k)}{(2\pi)^4} 
\theta(k_0)\theta(k^2) e^{-ikx} .
\end{equation}
We see that (\ref{2p2}) is compatible with (\ref{2p1}), provided that we take
\begin{equation}
F(k)=\sqrt{A_{d_{\cal U}} (2\pi)^4} \, (k^2)^{(d_{\cal U}-2)/2} .
\end{equation}
Thus, we see that the unparticle field operator introduced in \cite{georgi}
can be expanded as
\begin{eqnarray}\label{exp'}
\Phi(x) & = & A_{d_{\cal U}}^{1/2} \int \frac{d^4k}{\sqrt{(2\pi)^4}} 
(k^2)^{(d_{\cal U}-2)/2} \theta(k_0)\theta(k^2)
\nonumber \\
& & \times [a(k)e^{-ikx}+a^{\dagger}(k)e^{ikx}] .
\end{eqnarray}
Despite the name ``unparticle", the operators $a(k)$ and $a^{\dagger}(k)$
in (\ref{exp'}) 
have a well defined particle interpretation, as discussed in Sec.~\ref{SEC2}.

Now let us resolve an apparent conceptual paradox.
On one hand, the operators $a^{\dagger}(k)$ and $a(k)$ create and destruct
states with an {\em integer} number of particles, completely independent of the value of the scale dimension $d_{\cal U}$.
On the other hand, it is shown in \cite{georgi} that the unparticle stuff
manifests as a non-integer number $d_{\cal U}$ of particles.
Is there a contradiction between these two results? Not at all!
Namely, the result of \cite{georgi} refers to a non-integer number
of {\em invisible} particles, that is, to the case in which the unparticle
stuff is not observed. On the other hand, as explained in Sec.~\ref{SEC2}, 
when the unparticle stuff is observed, then it manifests as ordinary particles
with arbitrary mass.

\section{Remarks on canonical quantization}
\label{SEC4}

As we have explained, the commutation relation (\ref{e1'}) is derived
from the canonical commutation relation (\ref{e0}), or more precisely,
from a version of (\ref{e0}) in which $m$ is a continuous parameter.
Is it possible to derive (\ref{e1'}) in an alternative way, from an
unparticle canonical commutation relation of the form 
\begin{equation}\label{e0'}
[\Phi({\bf x}),\Pi({\bf x}')]=i\delta^3({\bf x}-{\bf x}') ,
\end{equation}
where $\Pi(x)$ is the unparticle canonical momentum with an 
expansion similar to (\ref{exp'})? Formally, it is possible to construct
an operator $\Pi(x)$ that obeys the commutation relation
(\ref{e0'}). Nevertheless, it is misleading to call such an operator
the ``canonical momentum". The very notion of the canonical momentum makes
sense only for Lagrangians that are functions of a canonical
coordinate and its first time derivative, but not other time derivatives.
On the other hand, the effective action describing unparticle fields 
is of the form \cite{kras,gaete}
\begin{equation}
S\propto \int d^4x \, \partial^\mu \Phi (-\partial^\nu\partial_\nu)^{1-d_{\cal U}}
\partial_\mu \Phi ,
\end{equation}
so the Lagrangian depends on other time derivatives of the unparticle
field $\Phi(x)$. Such systems do not admit the usual canonical quantization.
Consequently, the commutation relation (\ref{e1'}) cannot be derived from
a more fundamental canonical commutation relation of the form of
(\ref{e0'}). This suggests that the massive fields $\phi(x,m)$ from which
(\ref{e1'}) is derived 
play a more fundamental role than the effective unparticle field $\Phi(x)$.

As a conclusion, we end with the following remark.
Given the fact that the unparticle field admits a particle interpretation
just as ordinary fields do, 
the name ``unparticle" seems somewhat misleading.
Given our results above, perhaps the expression ``uncanonical field"
would better describe what this scale invariant field theory really is.

\section*{Acknowledgments}
This work was supported by the Ministry of Science of the
Republic of Croatia under Contract No.~098-0982930-2864.


\begin{thebibliography}{99}

\bibitem{georgi}
H. Georgi, {\it Phys.~Rev.~Lett.}~{\bf 98},  221601 (2007).
 
\bibitem{misc1}
K. Cheung, C.~S. Li and T.-C. Yuan, arXiv:0711.3361.
\bibitem{misc2}
Y.-F. Wu and D.-X. Zhang, arXiv:0712.3923.
\bibitem{misc3}
K. Cheung, T. W. Kephart, W.-Y. Kheung and T.-C. Yuan, 
{\it Phys. Lett. B} {\bf 662}, 436 (2008).

\bibitem{step}
M. A. Stephanov, {\it Phys.~Rev.~D} {\bf 76}, 035008 (2007).

\bibitem{kras}
N. V.~Krasnikov, {\it Int.~J.~Mod.~Phys.~A} {\bf 22}, 5117 (2007).

\bibitem{gaete}
P.~Gaete and E.~Spallucci, {\it Phys. Lett. B} {\bf 661}, 319 (2008).

\bibitem{kras2}
N. V.~Krasnikov, {\it Phys.~Lett.~B} {\bf 325}, 430 (1994).

\end{thebibliography}
\end{document}